\documentclass[3p,times]{elsarticle}

\usepackage{ecrc}
\usepackage{epsfig}

\volume{00}

\firstpage{1}

\journalname{Nuclear Physics A}

\runauth{}

\jid{npa}

\jnltitlelogo{Nuclear Physics A}

\usepackage{amssymb}

\biboptions{square,comma,numbers,sort&compress}

\usepackage[figuresright]{rotating}

\newcommand{\sqrtsnn}{$\sqrt{s_{NN}}$~}
\newcommand{\sqrts}{$\sqrt{s}$~}
\newcommand{\jpsi}{J/$\psi$~}
\newcommand{\jpsiwos}{J/$\psi$}
\newcommand{\psitwos}{$\psi\left(2S\right)$~}
\newcommand{\psitwoswos}{$\psi\left(2S\right)$}
\newcommand{\chic}{$\chi_c$~}
\newcommand{\upsi}{$\Upsilon\left(1S\right)$~}
\newcommand{\upsiwos}{$\Upsilon\left(1S\right)$}

\newcommand{\upsitwoswos}{$\Upsilon\left(2S\right)$}
\newcommand{\upsithrees}{$\Upsilon\left(3S\right)$~}
\newcommand{\chib}{$\chi_b$~}
\newcommand{\RpPb}{$R_{\rm pPb}$~}
\newcommand{\RAA}{$R_{\rm AA}$~}
\newcommand{\RAB}{$R_{\rm AB}$~}
\newcommand{\TAB}{$T_{\rm AB}$}
\newcommand{\pt}{$p_{\rm T}$~}
\newcommand{\ptwos}{$p_{\rm T}$}
\newcommand{\pPb}{\mbox{p-Pb}~}
\newcommand{\PbPb}{\mbox{Pb-Pb}~}

\newcommand{\pp}{pp~}
\newcommand{\ppwos}{pp}
\newcommand{\ccbar}{$c\bar{c}$~}
\newcommand{\meanptsquare}{$\langle p_T^2 \rangle$~}
\newcommand{\ptbroad}{$\Delta \langle p_T^2 \rangle = \langle p_T^2 \rangle_{pPb} - \langle p_T^2 \rangle_{pp}$~}
\newcommand{\deltaptsquare}{$\Delta \langle p_T^2 \rangle$~}

\begin{document}

\begin{frontmatter}



\dochead{}

\title{Quarkonium production in ALICE at the LHC}


\author{Cynthia Hadjidakis, for the ALICE Collaboration}

\address{Institut de Physique Nucl\'eaire d'Orsay, Universit\'e Paris-Sud, CNRS-IN2P3, Orsay, FRANCE}

\begin{abstract}
In heavy-ion collisions at the LHC, the ALICE Collaboration is studying Quantum Chromodynamics (QCD) matter at very high energy density where the formation of a Quark Gluon Plasma (QGP) is expected. Quarkonium production is an important probe to characterize the QGP properties. High precision data in pp collisions provide the baseline of $\mbox{Pb-Pb}$ measurements and $\mbox{p-Pb}$ collisions serve to quantify the amount of initial and/or final state effects, related to cold nuclear matter, that are largely unknown at the LHC energy. Since 2010, the LHC provided $\mbox{Pb-Pb}$ collisions at $\sqrt{s_{NN}}$ = 2.76 TeV, pp collisions at various energies and in 2013 $\mbox{p-Pb}$ collisions at $\sqrt{s_{NN}}$ = 5.02 TeV. In ALICE, quarkonia can be reconstructed at forward rapidity in the dimuon channel and at mid-rapidity in the dielectron channel, and, for both channels, down to zero transverse momentum. New measurements on inclusive production of $J/\psi$, $\psi (2S)$ and $\Upsilon$ performed in $\mbox{p-Pb}$ collisions and on the $p_{\rm T}$ dependence of inclusive $J/\psi$ in $\mbox{Pb-Pb}$ collisions are presented. The contribution of $J/\psi$ from $B$ hadrons to the inclusive production in $\mbox{Pb-Pb}$ is also discussed. Finally, an estimation of the cold nuclear matter effect in $\mbox{Pb-Pb}$, extrapolated from $\mbox{p-Pb}$ measurements, is given.
\end{abstract}


\end{frontmatter}


\section{Probing quark gluon plasma and cold nuclear matter with quarkonia}
\label{}
The ALICE collaboration studies heavy-ion collisions at the LHC in order to investigate nuclear matter at very high energy density where the formation of a QGP is expected. Quarkonium suppression was proposed as an important probe to characterize the QGP. It was predicted that at sufficiently high density medium, the color-screening of the heavy-quarks potential in deconfined QCD matter would lead to a sequential suppression of the quarkonium states production~\cite{Matsui-Satz:1986}. However, at large collision energy, where the large charm quark density may favour charmonium production by combination of charm quarks, charmonium enhancement was also anticipated~\cite{PBM-Stachel:2000,Thews:2000}.
Measurements carried out at SPS and RHIC revealed a suppression of the \jpsi production for the most central collisions. 
More recently at the LHC~\cite{ALICE-AA-jpsi:2013}, a suppression of \jpsi has also been measured with, however, a lower amplitude than the ones obtained at lower energies. In addition, the indication, for semi-central \PbPb collisions, of a non-zero \jpsi elliptic flow has been observed at low \ptwos~\cite{ALICE-AA-jpsi-v2:2013}. These results suggest a non-negligible contribution from \jpsi produced via the combination of charm quarks in or at the phase boundary of the QGP. For a correct interpretation of the energy, rapidity and \pt dependence of the measured suppression, different effects such as those due to cold or hot nuclear matter have to be considered.
 The cold nuclear matter (CNM) effects include initial and/or final state effects, such as nuclear shadowing, gluon saturation, energy loss and nuclear absorption that influence the quarkonium production without the need of the QGP formation. These CNM effects can be studied in \pPb collisions. 
Medium effects in proton(nucleus)-nucleus (A--B) collisions are measured with the nuclear modification factor \RAB defined as the invariant yield of a quarkonium state ($Y_{Q\bar{Q}}$) measured in A--B collisions normalized by the \pp cross-section scaled by the nuclear overlap function (\TAB) calculated using the Glauber model~\cite{glauber-alice:2013,Chiara:2013}. \RAB is expected to be equal to unity if the A--B collisions consist of an incoherent superposition of \pp collisions. 
The LHC provided \PbPb collisions at \sqrtsnn = 2.76 TeV in 2010 and 2011, and \pPb collisions at \sqrtsnn = 5.02 TeV in 2013. The large data sample allowed to study the production of various quarkonium states and, for the \jpsi case, differentially in \pt and $y$. In this proceeding, we will discuss preliminary measurements in \pPb collisions of \jpsi production at mid-rapidity (in the dielectron channel) and \jpsi, \psitwos and \upsi at forward rapidity (in the dimuon channel). Then we will present in \PbPb collisions, the \pt dependence of \RAA in particular at mid-rapidity and we will discuss the contribution from $B$ hadron decays to the inclusive measurement at mid-rapidity. Most of the measurements are inclusive and include in addition to the quarkonium direct production, contributions from the decay of higher mass excited states (\psitwos and \chic for the \jpsi and \upsitwoswos, \chib and \upsithrees for the \upsiwos). In the case of the \jpsi and \psitwoswos, the non-prompt production from the decay of $B$ mesons contributes as well. 

\section{Experimental apparatus and data sample}
The ALICE experiment~\cite{ALICE-detector:2008} has in the central region ($|\eta|<0.9$) detectors positioned in a large solenoidal magnet providing a uniform magnetic field of 0.5 T. The main tracking devices consist of the Inner Tracking System (ITS), made of six layers of silicon detectors that surround the beam pipe, and the Time Projection Chamber (TPC), a large cylindrical drift gas detector. In addition, the TPC provides particle identification via the measurement of the specific energy loss $dE/dx$. In the forward region ($-4<\eta<-2.5$\footnote{In the ALICE reference frame, the muon spectrometer covers a negative $\eta$ range}), the muon spectrometer consists of a front absorber of 10 interaction lengths ($\lambda_I$), a large 3 ${\rm T\cdot m}$ dipole magnet, a high granularity tracking system of ten detection planes and a muon filter wall (7.2 $\lambda_I$) followed by four planes of trigger chambers. 
Two forward VZERO scintillator hodoscopes ($2.8<\eta<5.1$ and $-3.7<\eta<-1.7$) are used for triggering and beam-induced background rejection. 
Finally, the Zero Degree Calorimeter (ZDC) placed 112.5 m from both sides of the Interaction Point allows to reject electromagnetic interactions and satellite collisions. 
Minimum Bias (MB) events were triggered by the two VZERO detectors. 
Dimuon events were triggered, in addition to the latter conditions, by two opposite-sign particles that fire the muon trigger system. To determine the centrality of a \PbPb collision, the VZERO amplitude was fitted by a geometrical Glauber-based model~\cite{glauber-alice:2013}.
Analysis on quarkonium measurements in \pPb and \PbPb collisions are described in more detail in~\cite{Michael:2013} for charmonia and in~\cite{Francesco:2013} for bottomonia. 

\section{Quarkonium measurements in \pPb collisions at \sqrtsnn = 5.02 TeV}

In \pPb collisions, due to the energy asymmetry of the LHC beams ($E_p$ = 4 TeV and $E_{Pb} = 1.58 \cdot A_{Pb}$ TeV, with $A_{Pb}$=208), the nucleon-nucleon center-of-mass system of the collisions is shifted by 0.465 in units of rapidity in the proton beam direction. Two beam configurations were provided by the LHC with the muon spectrometer in the Pb-going side or in the p-going side, defining two different rapidity ranges, backward or forward, respectively, for the dimuon analysis. The results presented here are based on a total integrated luminosity in \pPb collisions at \sqrtsnn = 5.02 TeV of 52 ${\rm \mu b}^{-1}$, 5 and 5.8 ${\rm nb}^{-1}$ for the mid ($-1.37<y_{cms}<0.43$), forward ($2.03<y_{cms}<3.53$) and backward ($-4.46<y_{cms}<-2.96$) rapidity intervals, respectively. The inclusive \jpsi production was measured in the dielectron (mid-rapidity) and dimuon channels (forward and backward rapidity) down to \pt = 0 GeV/c. In the dimuon channel, the inclusive \psitwos and \upsi were also measured. Since the \pp cross-sections were not measured at \sqrtsnn = 5.02 TeV, interpolations in energy and rapidity of the existing cross-sections were carried out~\cite{InterpolJpsiForw:2013,InterpolJpsiMid:2013,Michael:2013,Francesco:2013}.
\begin{figure}[hbt]
\begin{minipage}{8.0cm}
\epsfig{file=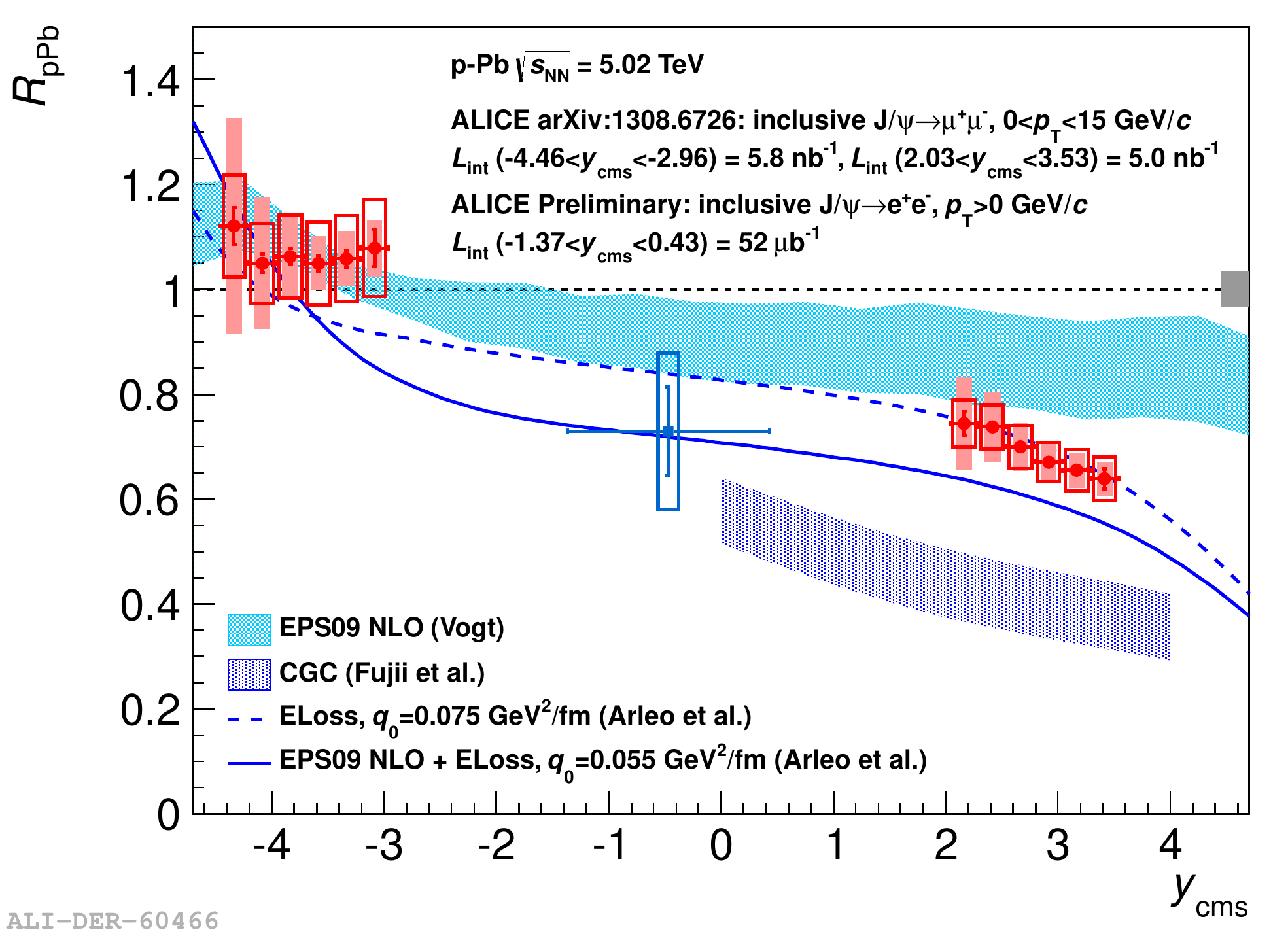,height=2.0in}
\end{minipage}
\begin{minipage}{8.0cm}
\epsfig{file=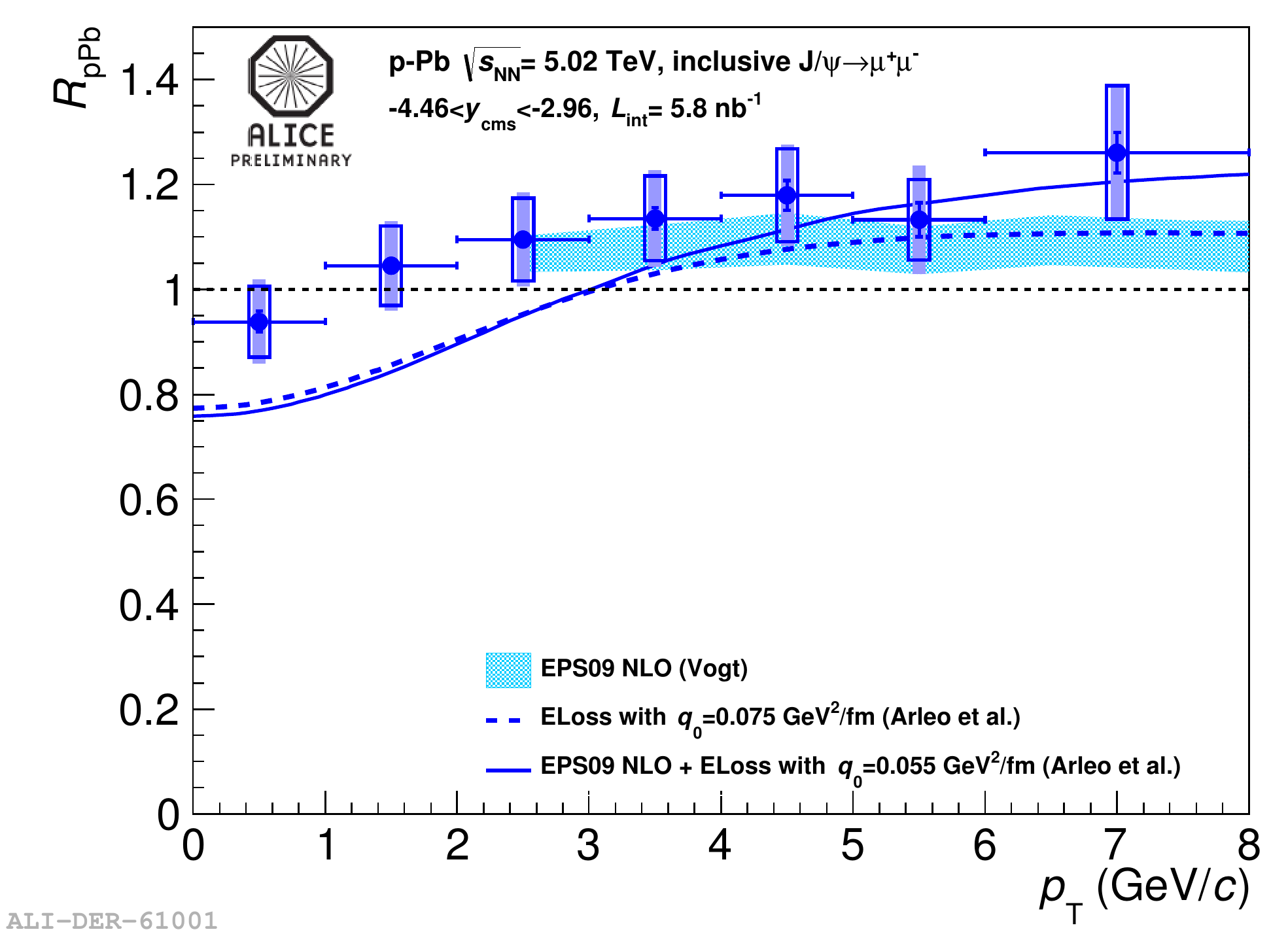,height=2.0in}
\end{minipage}
\begin{minipage}{8.0cm}
\epsfig{file=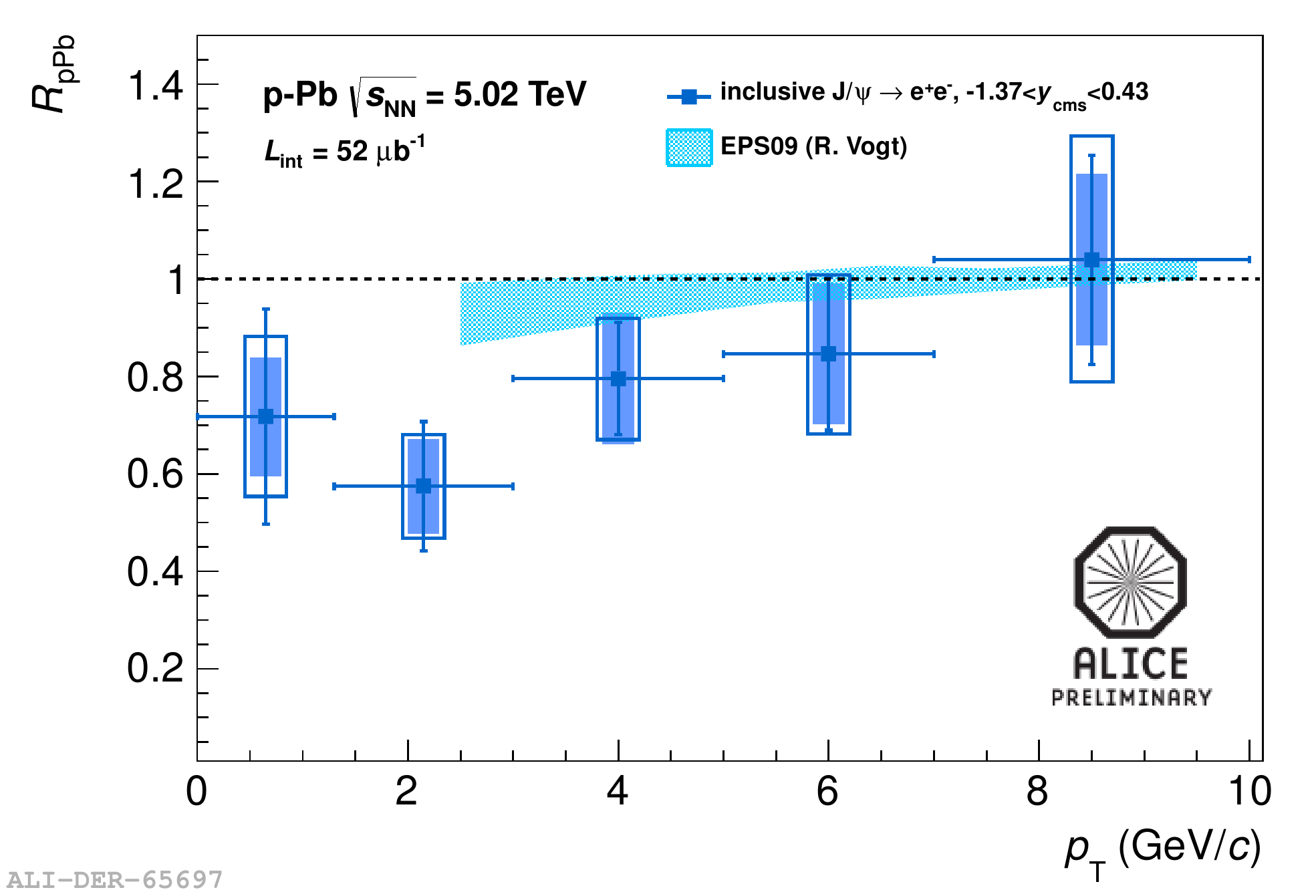,height=2.0in}
\end{minipage}
\begin{minipage}{8.0cm}
\epsfig{file=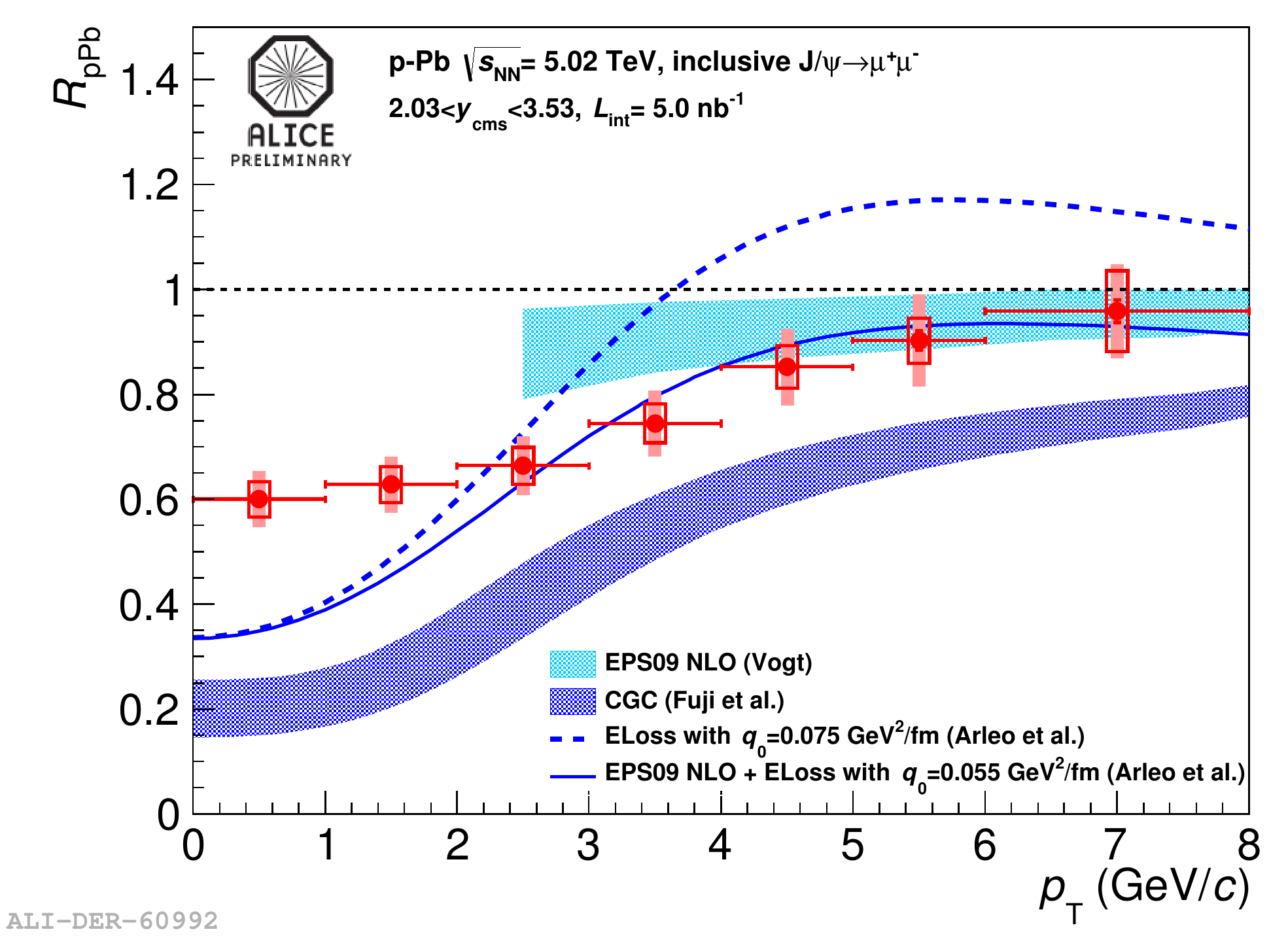,height=2.0in}
\end{minipage}
\caption{Inclusive \jpsi \RpPb as a function of rapidity (top left) and as function of \pt for three rapidity ranges (top right, bottom left and bottom right). See text for details.}
\label{fig:JpsiRpA}
\end{figure}
The nuclear modification factor \RpPb for inclusive \jpsi is presented in the top left panel of Fig.~\ref{fig:JpsiRpA} as a function of rapidity. Statistical uncertainties are shown as vertical error bars while the boxes, shaded area and box at unity represent the uncorrelated, partially and fully correlated systematic uncertainties, respectively. While the \jpsi production does not seem to be affected, within the uncertainties, by the nuclear environment at backward rapidity, the production is suppressed at mid and forward rapidity. Since the data refer to inclusive production, they include also non-prompt \jpsi produced by $B$ hadron decays. In pp at \sqrts = 7 TeV, the non-prompt \jpsi contributes to the inclusive cross-section by about 15\% for $|y|<0.9$~\cite{ALICEpp7prompt:2012} and 10\% for $2<y<4.5$~\cite{LHCbpp7:2011}. A non-prompt \jpsi \RpPb (integrated over \ptwos) was found by the LHCb experiment to be $0.83\pm0.08$ and $0.98\pm0.12$ for $2.5<y<4$ and $-4<y<-2.5$~\cite{LHCbpA:2013}, respectively. Assuming that \RpPb of non-prompt \jpsi varies from 0.6 to 1.3, its effect on the measured inclusive \RpPb was estimated to be at most 14\% (25\%) at low (high) \ptwos. 
Various models for prompt \jpsi production are compared to the data. First, a model based on nuclear parton distribution functions (nPDF) EPS09 associated to the Color Evaporation Model (CEM) at NLO~\cite{Vogt:2013} for the \jpsi production (shadowing model) reproduces well the rapidity dependence of the suppression. The theoretical uncertainties arise from the EPS09 nPDF and from the mass, the factorization and renormalization scale uncertainties on the cross-section calculation. At forward rapidity, the data favour a stronger shadowing than that predicted by this model. A second model includes medium-induced gluon radiation by the initial parton and the produced \ccbar pair and describes the nuclear suppression as due to the interference of the radiation before and after the hard production vertex~\cite{Arleo:2013,Arleo:2013pt} (coherent energy loss model). In this model, a parameterization of the \pp cross-section is used and the EPS09 nPDF are included or not. The data are better described with the coherent energy loss only. Finally a model assuming the gluon density saturation in the nucleus within the Color Glass Condensate (CGC) effective theory and using the CEM for the production of \jpsi~\cite{Fuji:2013} (CGC model) is also shown. The variation of the saturation scale and the charm quark mass defines the uncertainty. This model is only valid in a given rapidity range and underestimates the data. In Fig.~\ref{fig:JpsiRpA}, \RpPb is also shown as a function of \pt for the three rapidity intervals: backward (top right), mid (bottom left) and forward (bottom right). At backward rapidity, \RpPb shows a mild \pt dependence and is compatible with unity. At mid-rapidity, \RpPb tends to increase with \ptwos. 
Finally at forward rapidity \RpPb increases with \pt and is compatible with unity for \pt larger than 5 GeV/c. The shadowing model calculations provide numerical values for \pt larger than 2.5 GeV/c where they describe fairly well the data for the three rapidity ranges. The coherent energy loss model overestimates the suppression at forward rapidity for \pt below 2 GeV/c and the CGC calculations underestimate the data in the full \pt range.

\begin{figure}[hbt]
\begin{minipage}{8.0cm}
\epsfig{file=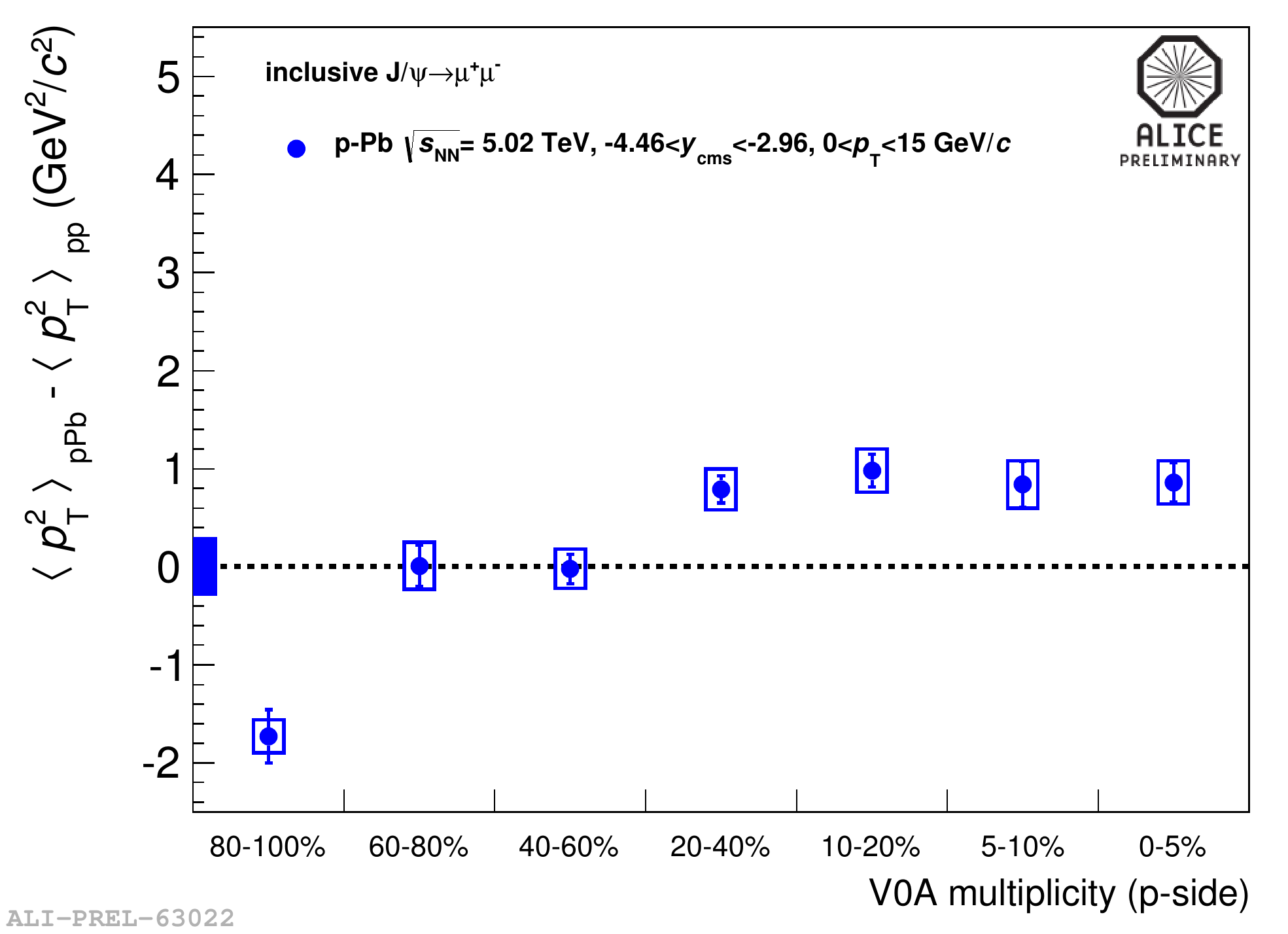,height=2.0in}
\end{minipage}
\begin{minipage}{8.0cm}
\epsfig{file=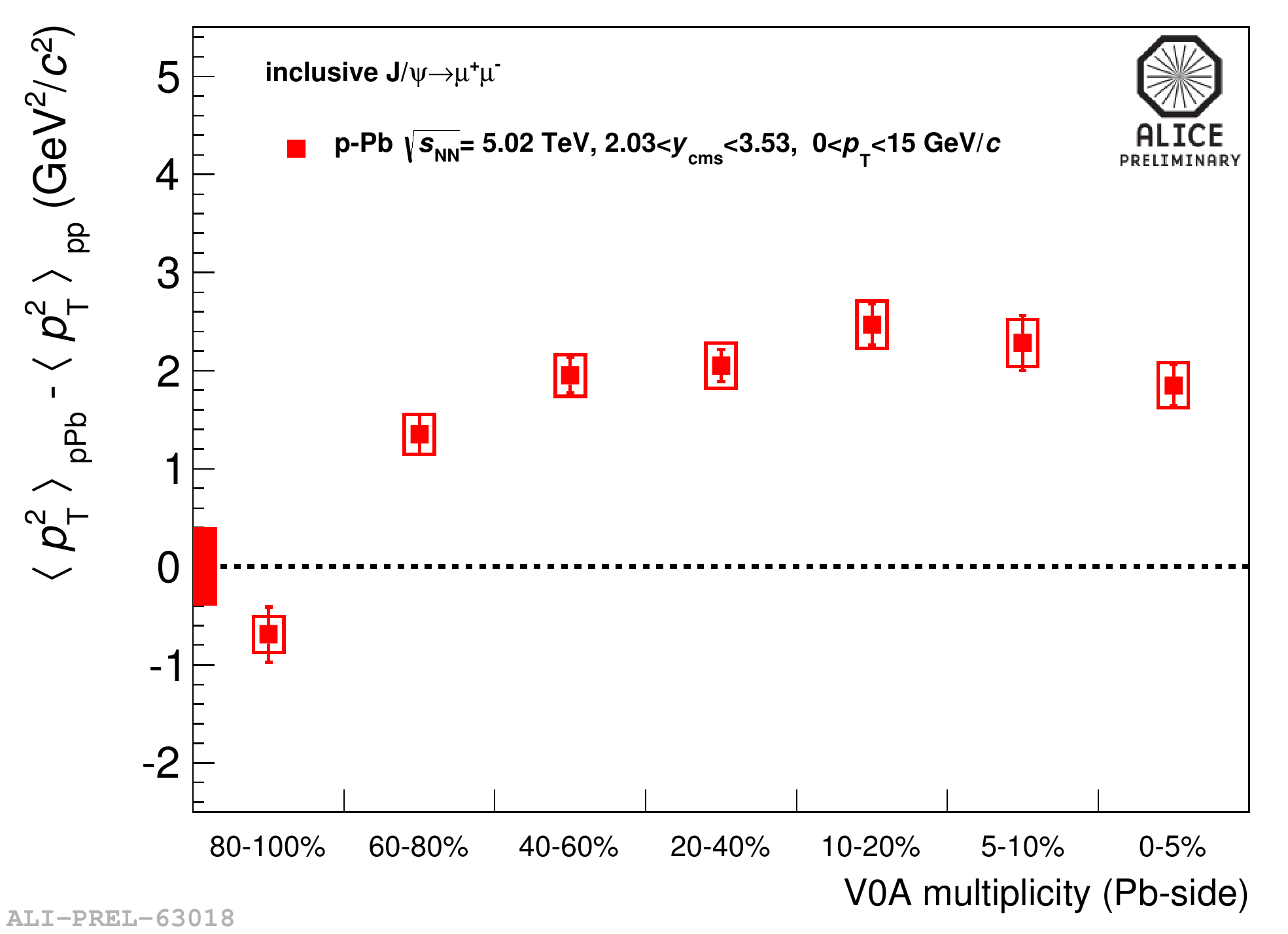,height=2.0in}
\end{minipage}
\caption{\ptbroad as a function of event multiplicity as measured by the VZERO detector at $-4.46<y_{cms}<-2.96$ (left panel) and at $2.03<y_{cms}<3.53$ (right panel). See text for details.}
\label{fig:meanptsquare}
\end{figure}

\begin{figure}[hbt]
\begin{minipage}{8.0cm}
\epsfig{file=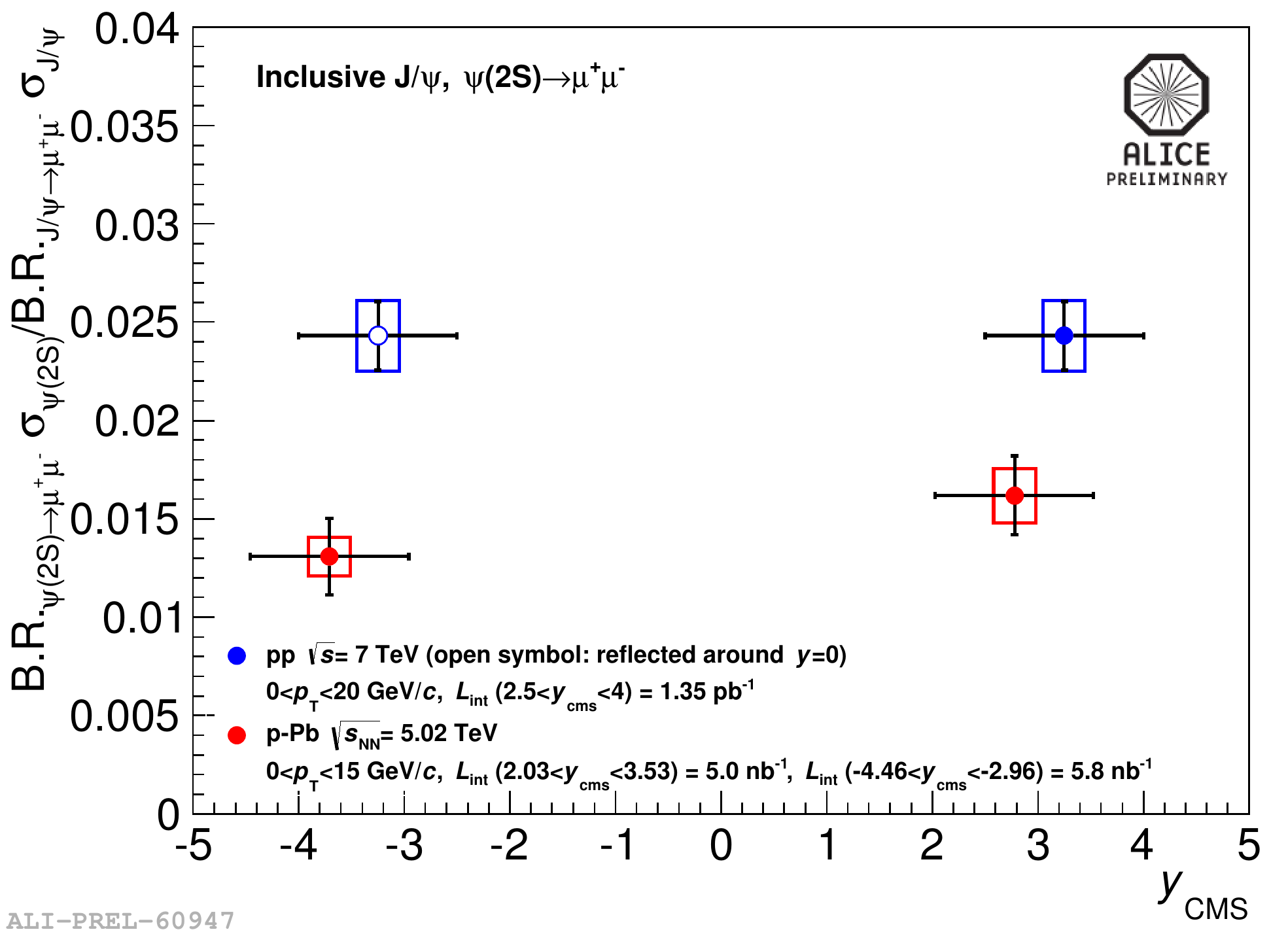,height=2.0in}
\end{minipage}
\begin{minipage}{8.0cm}
\epsfig{file=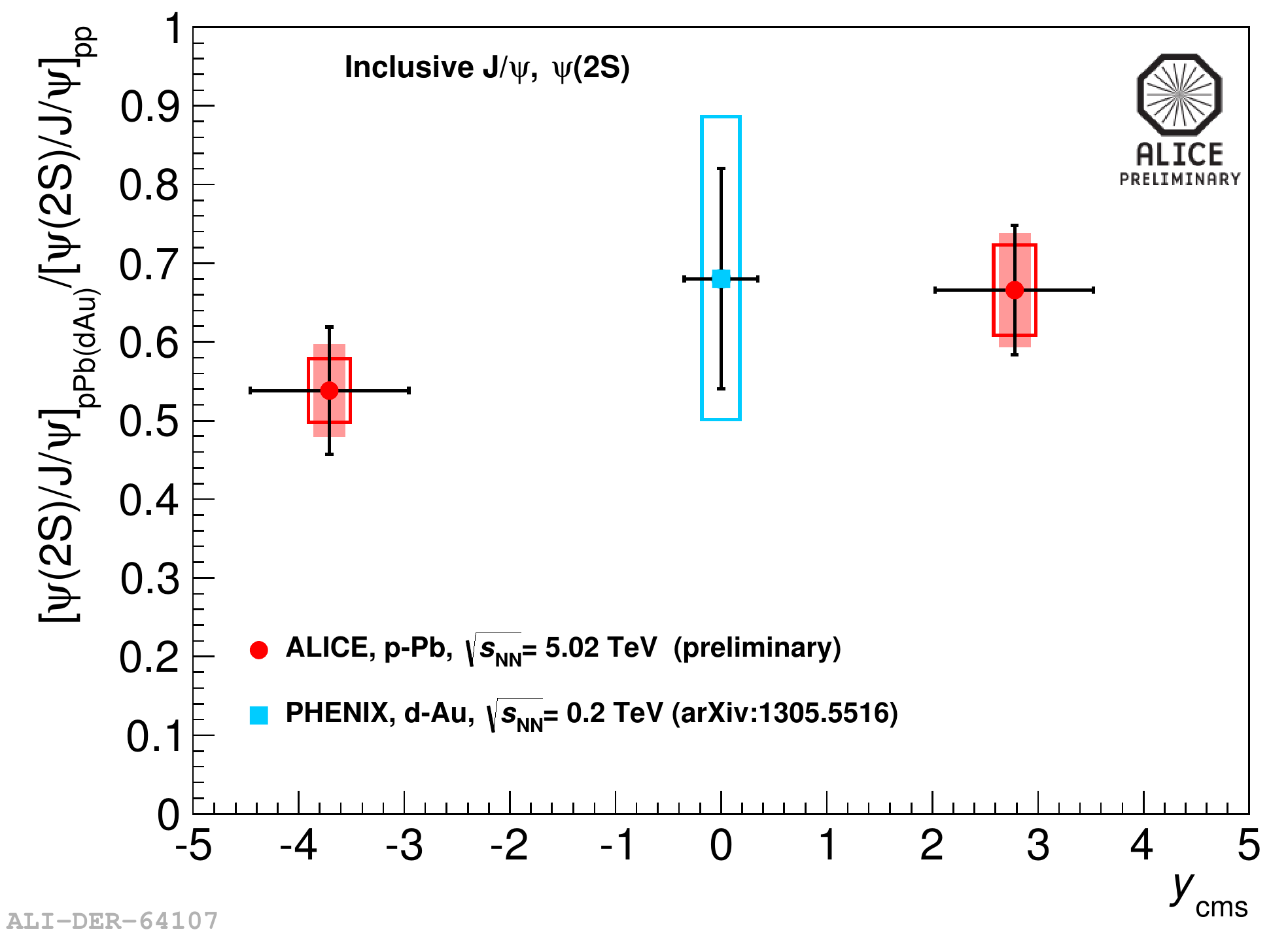,height=2.0in}
\end{minipage}
\caption{Left: inclusive \psitwos to \jpsi ratio in \pPb as compared to \ppwos. Right: double ratio of \psitwos to \jpsi in \pPb over that in \pp. The PHENIX result is also shown. See text for details.}
\label{fig:psi2S}
\end{figure}

In the dimuon analysis, the \pt dependence of the invariant yield was also studied as a function of the event multiplicity measured with an arm of the VZERO detector (V0A), located opposite to the muon spectrometer ($2.8<\eta<5.1$). 
Depending on the sense of the orbits of the beams, the V0A is located either in the p-going side (backward rapidity) or in the Pb-going side (forward rapidity). The \meanptsquare was then extracted from the \pt differential distributions and compared to the interpolated value in \pp collisions. The \pt broadening \ptbroad is shown in Fig.~\ref{fig:meanptsquare} as a function of the V0A event multiplicity. \deltaptsquare is larger at forward than at backward rapidity and increases in both cases as a function of the event multiplicity. 
In the dimuon analysis, the \psitwos yield integrated over \pt was extracted for the backward and forward rapidity intervals. The inclusive \psitwos to \jpsi ratio in \pPb at \sqrtsnn = 5.02 TeV is shown in Fig.~\ref{fig:psi2S} and is compared to the same ratio measured in \pp at \sqrts = 7 TeV. Models such as the shadowing, coherent energy loss or CGC-based model presented earlier do not predict different nuclear effects for the \psitwos and the \jpsi, hence they predict this ratio to be the same in \pp and \pPb collisions. The data show instead that the \psitwos is clearly more suppressed than the \jpsi both at backward and forward rapidity, pointing out to the presence of final state effect to explain this observation. The double ratio of \psitwos to \jpsi in \pPb as compared to \pp at \sqrtsnn = 5.02 TeV was also derived. Here additional systematic uncertainties arise from the interpolation in energy and rapidity of the \psitwos to \jpsi ratio in \pp. The relative suppression of the \psitwos as compared to the \jpsi was already observed at mid-rapidity in d--Au collisions at \sqrtsnn = 200 GeV~\cite{PHENIXpsi2s:2013} and is now confirmed with larger significance by ALICE in \pPb collisions at \sqrtsnn = 5.02 TeV.
Finally, the inclusive \upsi was measured at backward and forward rapidity in the dimuon channel and the \RpPb is shown in the left panel of Fig.~\ref{fig:UpsandRaajpsi}. It is compared to a shadowing model~\cite{Ferreiro:2013} (EPS09 LO nPDF associated with a Color Singlet Model at LO). The \upsi has a similar suppression at forward rapidity than the \jpsi and seems more suppressed than predicted by the two shadowing models mentioned before~\cite{Vogt:2013,Ferreiro:2013} as well as by the coherent energy loss model~\cite{Arleo:2013} (not shown here). However the large uncertainties dominated by the fully correlated uncertainties from the energy interpolation of the pp cross-section do not allow to constrain the models. 

\section{Quarkonium measurements in \PbPb collisions at \sqrtsnn = 2.76 TeV}

\begin{figure}[hbt]
\begin{minipage}{8.0cm}
\epsfig{file=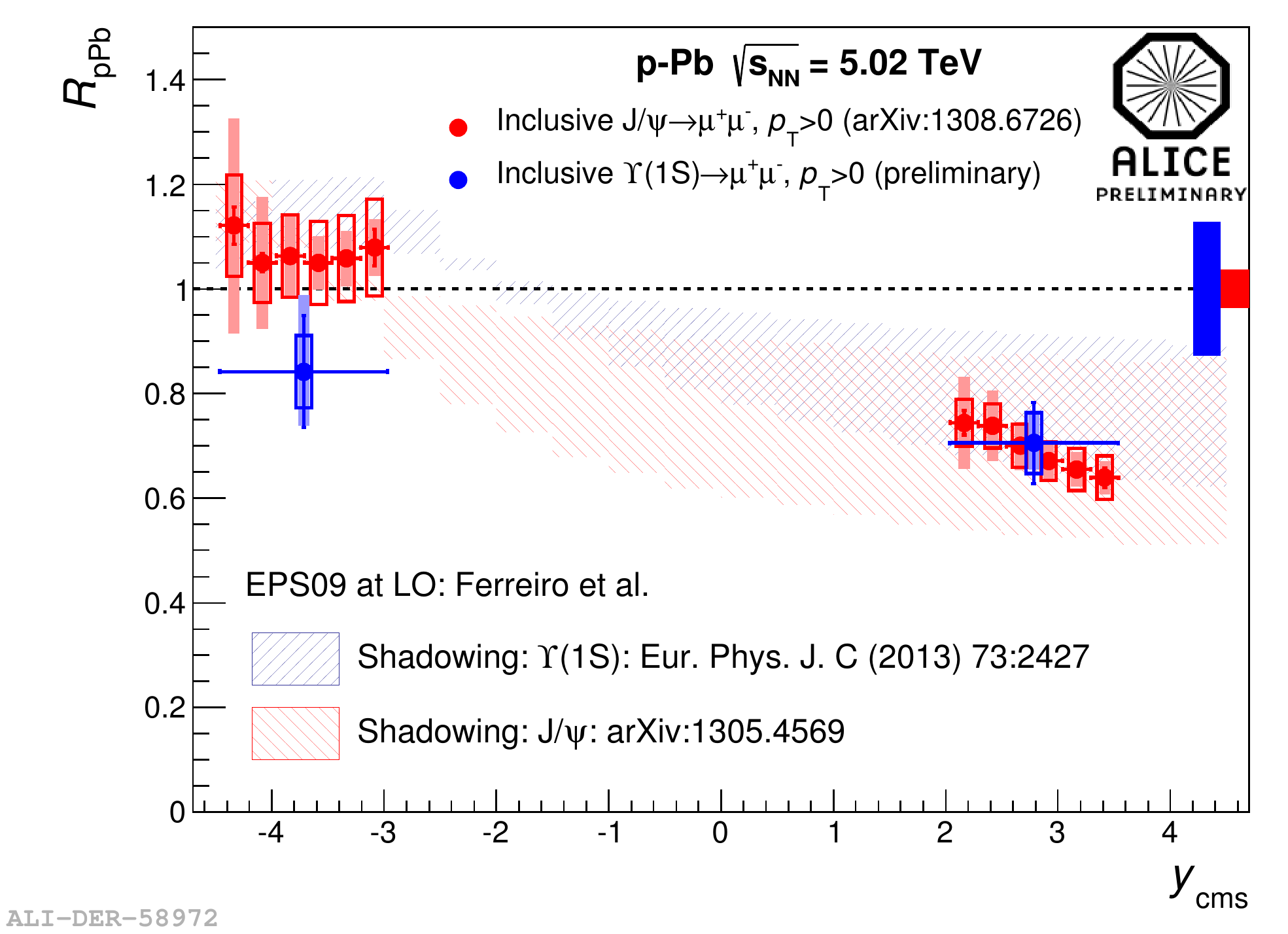,height=2.0in}
\end{minipage}
\begin{minipage}{8.0cm}
\epsfig{file=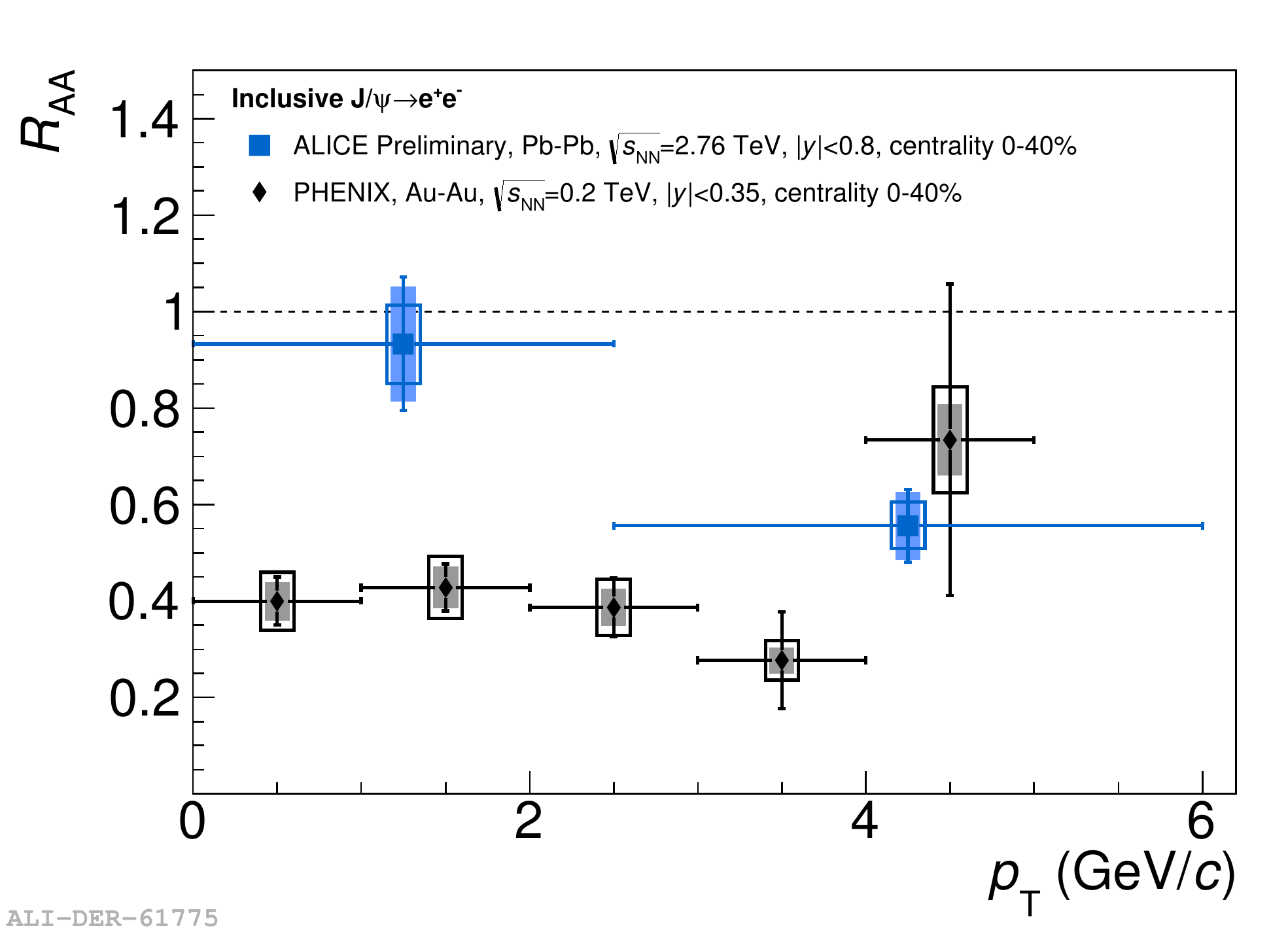,height=2.0in}
\end{minipage}
\caption{Left: inclusive \upsi \RpPb as a function of rapidity. Right: inclusive \jpsi \RAA as a function of \pt at mid-rapidity. See text for details.}
\label{fig:UpsandRaajpsi}
\end{figure}

\begin{figure}[hbt]
\begin{minipage}{8.0cm}
\epsfig{file=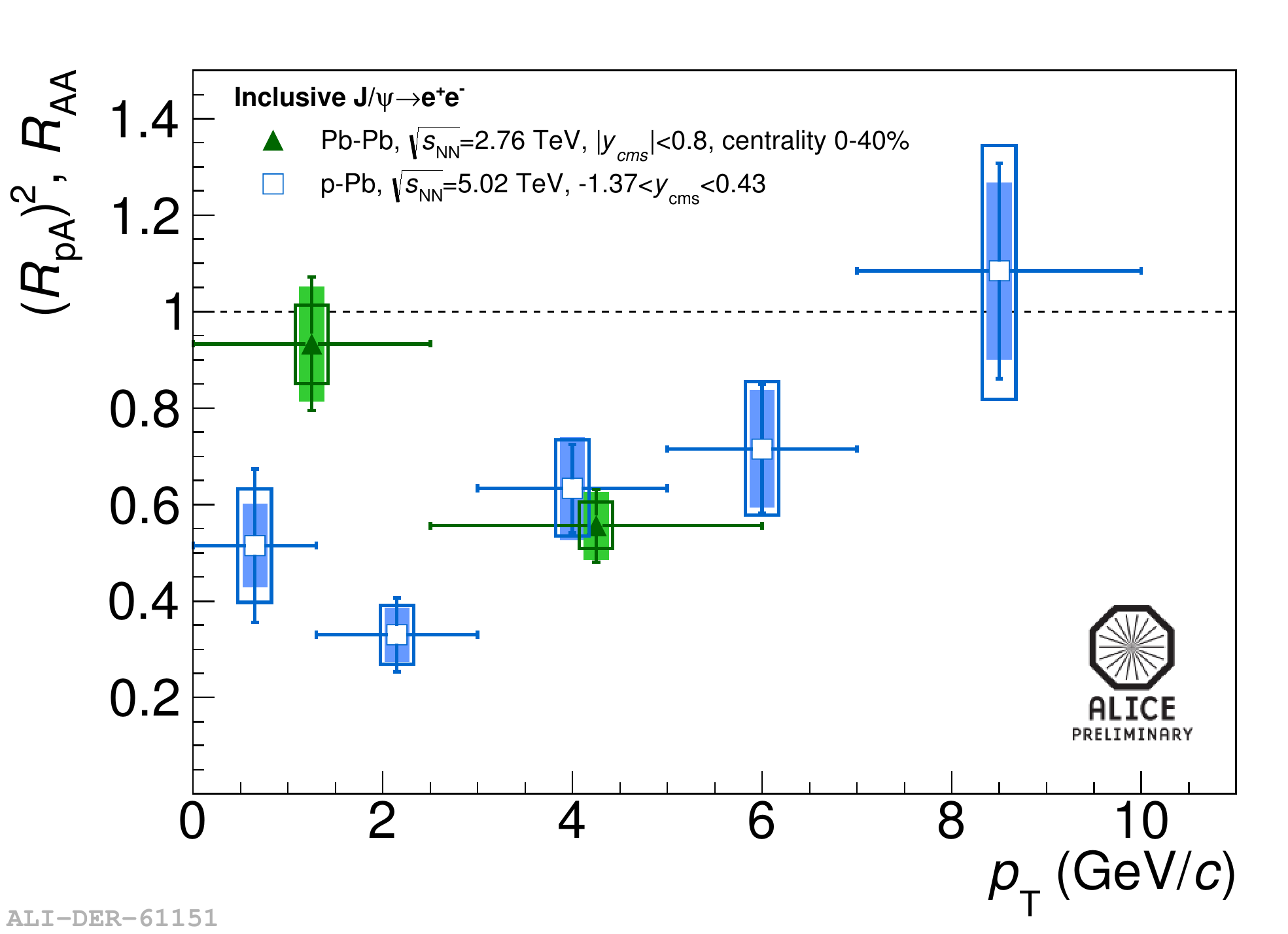,height=2.0in}
\end{minipage}
\begin{minipage}{8.0cm}
\epsfig{file=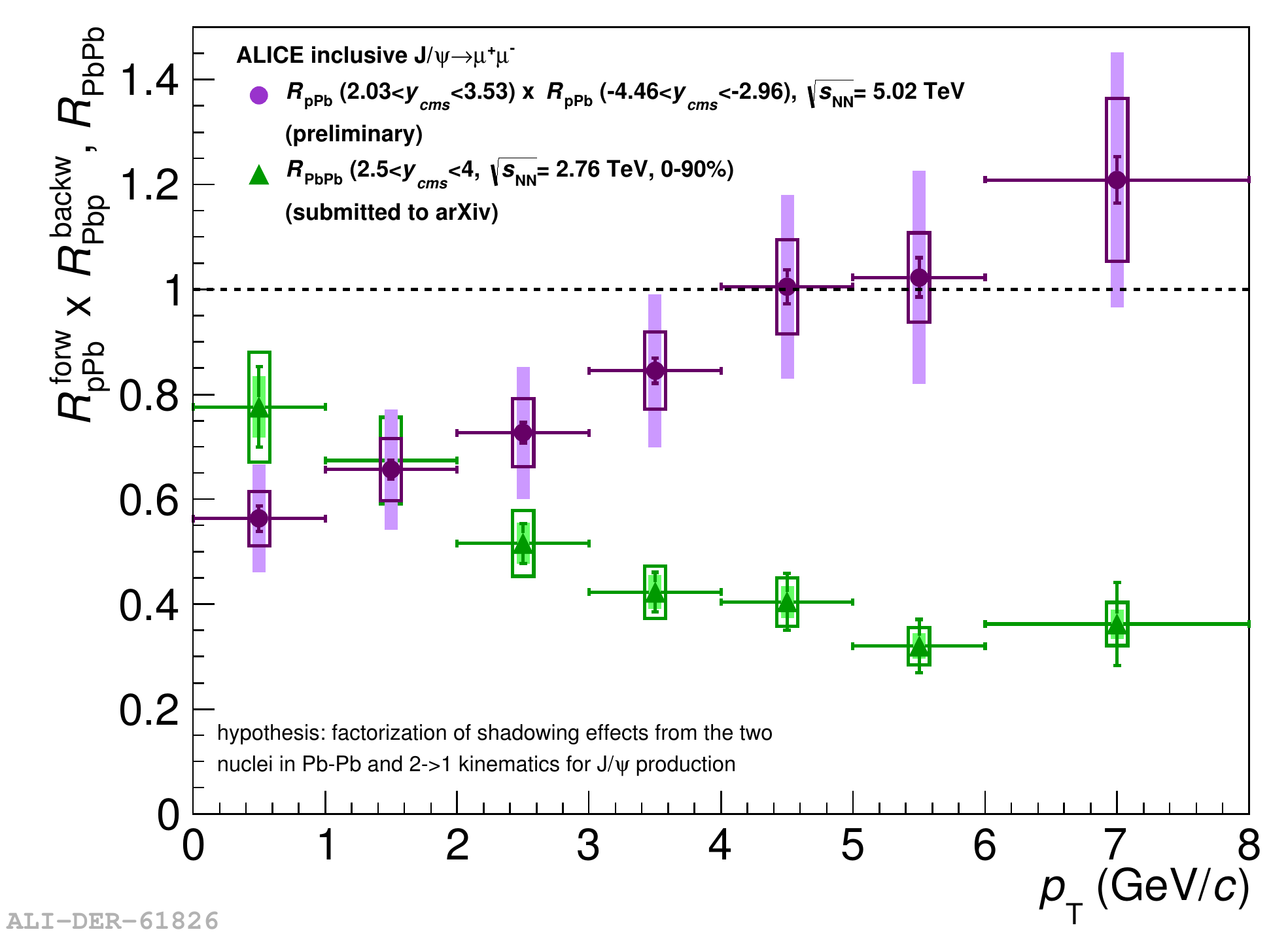,height=2.0in}
\end{minipage}
\caption{Inclusive \jpsi \RAA at mid (left panel) and forward (right panel) rapidity as a function of \ptwos. An estimation of shadowing effect in \PbPb measured with \jpsi in \pPb collisions is also displayed. See text for details.}
\label{fig:Raajpsi}
\end{figure}

The results presented here are based on a total integrated luminosity in \PbPb collisions at \sqrtsnn = 2.76 TeV of 28 and 69 ${\rm \mu b}^{-1}$ for the mid ($|y|<0.8$) and forward ($2.5<y<4$) rapidity intervals, respectively. The inclusive \jpsi production was measured in the dielectron (mid-rapidity) and dimuon channels (forward rapidity) down to zero \ptwos. The centrality dependence was also measured for the \jpsiwos~\cite{ALICE-AA-jpsi:2013} and, at forward rapidity, for the \psitwos to \jpsi ratio~\cite{ALICEQM2012:2012} and \upsiwos~\cite{Manceau:2013}. The \pp cross-section was measured at \sqrts = 2.76 TeV~\cite{ALICEpp2p76:2012} and is used for the estimation of \RAA at forward rapidity. At mid-rapidity an interpolation procedure is preferred since the resulting uncertainty is lower and the interpolated cross-section is in agreement with the measurement. 
The right panel of Fig.~\ref{fig:UpsandRaajpsi} shows \RAA as a function of \pt at mid-rapidity for the events corresponding to the most central $40\%$ of the \PbPb inelastic cross-sections. The \jpsi is modestly or not suppressed at \pt $\approx 1$ GeV/c while a suppression of $\approx40\%$ is found at \pt $\approx 4$ GeV/c. The fraction of non-prompt \jpsi to inclusive \jpsi was measured to be $0.133\pm 0.043(stat)\pm 0.013(syst)$ at $|y|<0.9$ for $2<p_{\rm T}<10$ GeV/c, a value close to the one measured in \pp collisions at \sqrts = 7 TeV~\cite{ALICEpp7prompt:2012}. This indicates that the contribution of \jpsi from $B$ hadrons has a small effect on the inclusive \RAA at low $p_{T}$. The measurements of prompt and non-prompt \jpsi \RAA are ongoing. The inclusive \jpsi \RAA is compared to PHENIX~\cite{PHENIXAuAuJpsi:2006} in the same event centrality range. The different \pt dependence between the two center-of-mass energies already reported at forward rapidity in ALICE~\cite{ALICE-AA-jpsi:2013} is also confirmed at mid-rapidity.
In order to quantify the effect from CNM for the /jpsi production in \PbPb collisions, the \pPb measurements were exploited with the following assumptions. First, the production mechanism for the \jpsi was assumed to be $g+g\rightarrow J/\psi$, where the \jpsi kinematics defines entirely the nucleon longitudinal momentum fractions, $x_{1,2}$, carried by the two initial gluons. This assumption allows us to compare the gluon $x$ in the nucleus in \pPb collisions at \sqrtsnn = 5.02 TeV and in \PbPb collisions at 2.76 TeV. They were found to be approximately similar despite different energies and rapidity domains. Second, it was assumed that CNM effects factorize in \pPb and in \PbPb as it is the case in the shadowing model. The influence of CNM effect on the nuclear modification factor in \PbPb can then be evaluated from $R_{AA}^{{\rm Shad}}=\left(R_{pPb}\right)^2$ at mid-rapidity and from $R_{AA}^{{\rm Shad}}=R_{pPb}\left(2.03<y_{cms}<3.53\right)\times R_{pPb}\left(-4.46<y_{cms}<-2.96\right)$ at forward rapidity ($\rm Shad$ refers to shadowing). The \pt dependence of the inclusive \jpsi \RAA is shown on Fig.~\ref{fig:Raajpsi} at mid (left panel) and forward rapidity (right panel). One should note that at mid and forward rapidity, \RAA is obtained for the $40\%$ and $90\%$ most central events, respectively. 
In Fig.~\ref{fig:Raajpsi} the nuclear modification factor \RAA are compared to the extrapolated CNM effect $R_{AA}^{{\rm Shad}}$. While for \pt above 7 (4) GeV/c at mid (forward) rapidity, the extrapolated CNM effect is small, at lower \pt the suppression observed in \PbPb can be ascribed only to this effect. 

\section{Conclusion}
Measurements of quarkonium production have been presented in \pPb collisions at \sqrtsnn=5.02 TeV and in \PbPb collisions at \sqrtsnn=2.76 TeV. 
In \pPb collisions the nuclear modification factor was measured for inclusive \jpsi as a function of rapidity and as a function of \pt for three different rapidity ranges. The measurements support a strong shadowing at forward rapidity and are in reasonable agreement with theoretical models based on nuclear shadowing with or without a contribution from coherent energy loss. The \jpsi \pt broadening \ptbroad was also studied and an increase with the event multiplicity was measured. The \psitwos was found to be more suppressed relatively to the \jpsi for both forward and backward rapidity and this behaviour is not explained by the models discussed in this paper~\cite{Vogt:2013,Ferreiro:2013,Fuji:2013,Arleo:2013}.
 For inclusive \upsiwos, the measurements show a similar suppression than for the \jpsi, with, however, large uncertainties that do not allow to constrain the models. 
In \PbPb collisions the \pt dependence of the nuclear modification factor was measured at mid-rapidity and similar conclusion as for the published results at forward rapidity~\cite{ALICE-AA-jpsi:2013} were found: the \jpsi is less suppressed at low \pt than at large \pt and this behaviour is clearly different from low energy data. In addition to the indication of a non-zero \jpsi elliptic flow in semi-central collisions~\cite{ALICE-AA-jpsi-v2:2013}, these results suggest a non-negligible contribution from \jpsi produced via combination of charm quarks in the QGP or at the phase boundary. Finally, the contribution of CNM effects in \PbPb was extrapolated from the \pPb measurement. Under the assumptions described in this paper, \RAA at low \pt would be, once corrected for CNM effects, even higher than the measured one reinforcing the interpretation of a recombination mechanism.





\bibliographystyle{elsarticle-num}
\bibliography{HP2013-Cynthia-biblio}







\end{document}